# Labyrinthine Island Growth during Pd/Ru(0001) Heteroepitaxy

N. Rougemaille,[1,2,3] F. El Gabaly,[1,4] R. Stumpf,[2] A. K. Schmid,[1] K. Thürmer,[2] N. C. Bartelt,[2] and J. de la Figuera[4,5]

[1]Lawrence Berkeley National Laboratory, Berkeley, California 94720, USA
[2]Sandia National Laboratories, Livermore, California 94550, USA
[3]Institut Néel, CNRS and Université Joseph Fourier, BP 166, F-38042 Grenoble Cedex 9, France
[4]Universidad Autónoma de Madrid, Madrid 28049, Spain
[5]Instituto de Química-Física "Rocasolano", CSIC, Madrid 28006 Spain

## Abstract

Using low energy electron microscopy we observe that Pd deposited on Ru only attaches to small sections of the atomic step edges surrounding Pd islands. This causes a novel epitaxial growth mode in which islands advance in a snakelike motion, giving rise to labyrinthine patterns. Based on density functional theory together with scanning tunneling microscopy and low energy electron microscopy we propose that this growth mode is caused by a surface alloy forming around growing islands. This alloy gradually reduces step attachment rates, resulting in an instability that favors adatom attachment at fast advancing step sections.

At sufficiently high temperature, epitaxial growth is expected to occur by the flow of atomic steps: atoms deposited on flat terraces on the surface of a crystal diffuse across the surface until they encounter steps and are incorporated at low energy kink sites, causing the steps to advance. Since the utility of thin films grown in this way often depends on the surface step structure, much work has been devoted to characterizing the morphology of steps. For example, if diffusion along the step edges is fast step edges become smooth, whereas slow step diffusion produces dendritic step edges [1]. In single component systems the basic atomic processes governing the morphology of moving steps are increasingly well understood [2]. However, in heteroepitaxy, new types of cooperative behavior can arise [3-5]. In this work we report the discovery of a novel way in which surface alloying and step motion couple, leading to a distinctive labyrinthine island morphology during growth in which only fast moving step edges are good sinks for deposited atoms. By unraveling the cooperative atomic scale processes leading to these striking 100-nm-scale patterns we make an inroad into understanding the intricate connections between surface morphology and alloying.

We use the complementary capabilities of real-time low energy electron microscopy (LEEM) and atomic resolution scanning tunneling microscopy (STM) to study the evolution and microscopic structure of Pd deposited on the Ru(0001) surface. Pd and Ru are somewhat soluble in each other (the maximum solubilities are a few percent at 900 K [6]) and thus at high enough temperature alloying during growth might be expected. We grow submonolayers of Pd by physical vapor deposition onto well cleaned Ru(0001) substrates at 840 K at base pressures below $1\times10^{-10}$ torr. Previous diffraction studies have shown that monolayer (ML) films of Pd on Ru are pseudomorphic [7]. A sequence of LEEM images [8] of the growing surface during Pd deposition at a substrate temperature of 840 K is shown in Fig. 1. Initially Pd starts to decorate the Ru substrate steps, forming a uniform narrow band at step edges. However this uniform step flow quickly stops and instead growth continues only from a few widely separated locations leading to a snakelike motion of islands, as shown in Fig. 1(a). These islands have an (approximately) constant width but grow longer as more Pd is deposited [see Figs. 1(b)-1(d)]. The direction of advance of the active end of an island occasionally changes, apparently randomly. The islands avoid connecting with themselves, other islands, and substrate steps - when the advancing active region comes close to an existing, inactive Pd layer, it turns away, and often splits in two [Figs. 1(b)-1(d)]. This entire process leads to a labyrinthine structure of the submonolayer Pd.

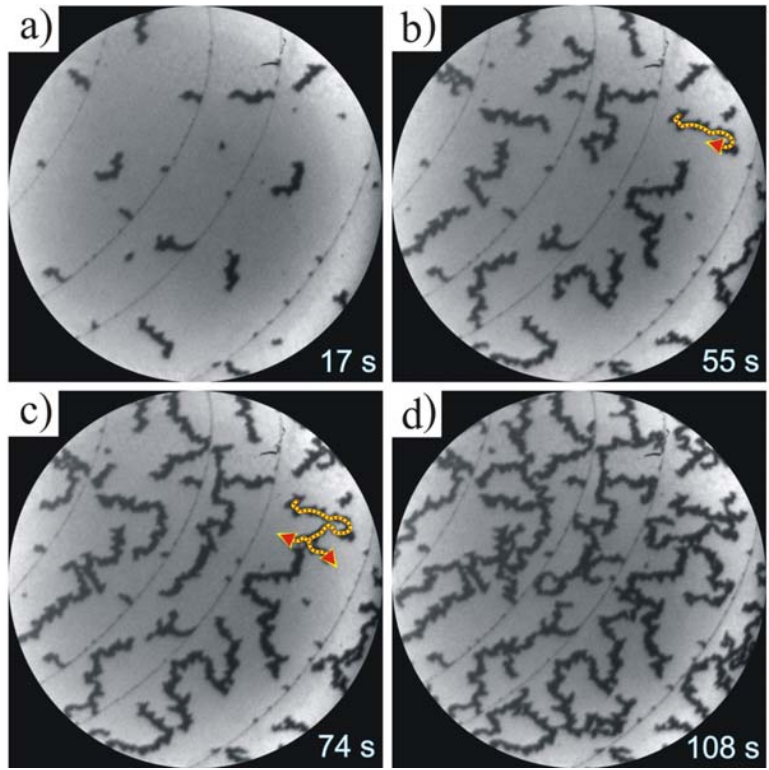

FIG. 1. Sequence of LEEM images of the Pd growth process on Ru(0001) at 840 K. Field of view is 6 μm. The monoatomic Ru steps (the four parallel lines that cross the images) are separated by approximately 1.5 μm. The deposition rate is 1 ML in 421 s. After first uniformly decorating the step edges in (a), the Pd grows only from several distinct locations. In (b) and (c) the self-avoiding nature of different growth fronts is evident: Colliding growth fronts turn aside or split rather than coalesce.

The observed growth mode is puzzling at first. All step edges should be good sinks for deposited Pd atoms, yet Pd is incorporated overwhelmingly at a few special regions of the step edge, causing them to advance. Also, growing islands should be easily able to coalesce. What atomic properties cause this growth behavior? One hypothesis is that the narrow islands are an example of a stress domain pattern; i.e., a narrow island width is chosen as a compromise between the energetic cost of creating step edges and the relaxation of surface stress at the island boundaries. For example, Pb overlayer islands grown on Cu(111) can exhibit a similar morphology [9], which has been characterized quantitatively as a stress domain pattern in thermodynamic equilibrium. We will show, however, that an entirely different, kinetic, mechanism is at work here.

To address the reasons for the snakelike growth we imaged the atomic structure of the surface with STM. We grew the Pd islands under the same conditions as in the LEEM experiments and then rapidly quenched the surface to room temperature before imaging it with STM. As seen in Fig. 2(a), the same meandering island shape was observed. Atomic resolution images with chemical contrast [10] show that there is a distinct difference between the atomic structure of the Ru terraces adjacent to steps that have apparently moved fast and those that have not: the terraces adjacent to the stationary step edges are heavily alloyed with Pd, with densities up to about 10%. On the other hand, very little alloying is observed adjacent to active, moving Pd step edges. These data immediately suggest that alloyed Ru regions have slowed the attachment of Pd adatoms on the Ru terraces to the edges of the Pd islands.

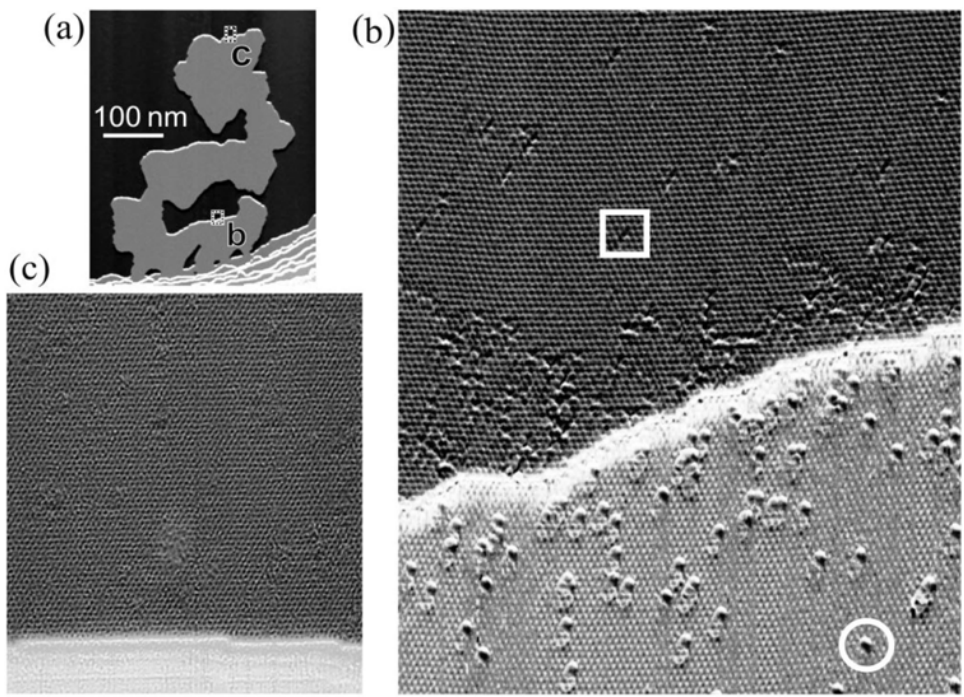

FIG. 2. STM images after deposition of 0.2 ML of Pd. (a) View of an entire Pd island. (b),(c) Atomic resolution images of particular regions of the islands shown in (a). Close to the step section (b) the Pd island is alloyed with Ru atoms (circled) and the Ru terrace is alloyed with Pd atoms (one is marked by a square). The degree of alloying is much lower at the active step section (c).

This suggestion raises two questions: By what mechanism does the alloying impede Pd incorporation at steps, and why is the alloying restricted to the region adjacent to the stationary island edge? To answer these questions we performed density functional theory (DFT) calculations [11] within the Perdew-Burke-Ernzerhof [12] generalized gradient approximation of the energetics of Pd on Ru(0001) using the VASP code [13].

First, we explain why alloying is rare on terraces far away from Pd step edges. Consider the energetics of the process labeled "A" in Fig. 3(a). The calculated energy difference between dispersed Pd substitutional atoms on the Ru terraces and Pd atoms in the overlayer is $E_f = 0.24$ eV. At the experimental temperature of 840 K this formation energy would give an equilibrium alloy density of approximately $\exp(-E_f/kT) = 0.04$. However, the lowest energy barrier we could find for a Pd adatom to exchange with a Ru atom in the top substrate layer [process "C" in Fig. 3(a)] is 2.48 eV. The resulting substitutional Pd and Ru adatom is 1.01 eV higher in energy than the Pd adatom initial state. Consequently, it is likely that the Ru adatom reexchanges with a substitutional Pd, because of the 1.01 eV energy gain and a barrier that is only 2.48 eV - 1.01 eV = 1.47 eV. The reexchange is avoided only if the Ru adatom is trapped at a step instead, where it ultimately gains the Ru adatom formation energy of 1.38 eV [process "B" in Fig. 3(a)]. Thus the alloying of Pd into the Ru surface by the adatom process is much more likely near steps than far away [14]. Consistent with this picture is the observation by STM of Ru atoms embedded in the Pd overlayer [Fig. 2(b)] with a density that parallels the density of alloyed Pd on the nearby Ru terrace. Furthermore, the density of Pd in the Ru terraces adjacent to the step edges is about 5%, on the order of what one would expect on the basis of the DFT formation energy.

We identify two mechanisms by which the existence of a step edge alloy can hinder the attachment of Pd adatoms at Pd step edges. First, the alloy on the terraces might impede the diffusion of Pd toward steps. Supporting this scenario, our calculations show a repulsion between Pd adatoms and Pd atoms substituted into the Ru substrate: only three of the six threefold hollow sites adjacent to an embedded Pd are local minima with a binding energy that is reduced by 0.14 eV over clean Ru(0001). This repulsion impedes diffusion of Pd near Pd islands. The alloy concentration $\rho$ near the step edges observed in STM is, at most, 10%. Our calculations [15] for this density give a decrease in the diffusion coefficient D by a factor of 2. If the alloying region is only a few nm wide as observed in STM, this decrease in D is by itself not sufficient to effectively block the step edges.

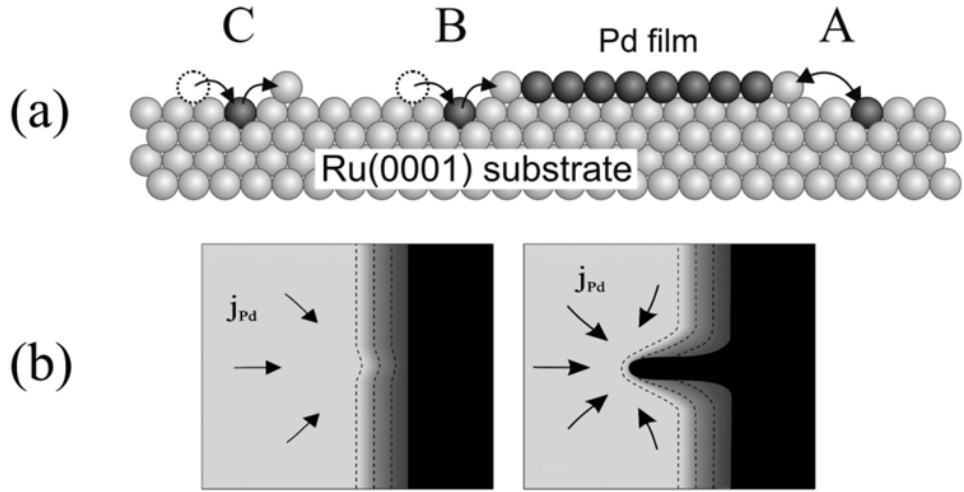

FIG. 3. (a) Various processes involved in the PdRu alloy formation. (b) Schematic of the instability of Pd step attachment caused by variations in alloy concentration in front of the step edges. The dotted lines represent lines of equal concentration of Pd alloyed into the Ru terraces. Regions with locally less alloy (left) become better sinks of Pd, leading to larger growth rates (right) and a further decrease in alloy concentration.

A second alloying effect that slows the Pd step advancement is that covering alloyed Pd atoms by the advancing Pd layer costs energy compared to advancing over the pure Ru substrate. The energy penalty is 0.5 eV per embedded Pd, almost equal to the typical binding energy of Pd to the Pd step of 0.61 eV. Given the low Pd adatom diffusion barrier of 0.19 eV it is reasonable to expect that local equilibrium along steps can be established: Pd adatoms that attach to a Pd island step located over a heavily alloyed substrate region are more likely to detach because of their higher energy; i.e., the Pd adatom density in equilibrium with a step edge section advancing over an alloyed terrace would be higher than over a clean Ru terrace. This would cause concentration gradients on the terraces leading to a Pd flux away from alloyed step edges and toward the edges that are in clean regions. Thus growth of Pd over alloyed regions will be avoided. Notice that the higher Pd adatom concentration produced by this mechanism would reinforce the mechanism discussed above by promoting the alloying of Pd atoms into the Ru substrate.

In any case, the degree of alloying near a stationary step edge, and hence the barrier to Pd attachment, depends on time. We checked experimentally the time dependence of the Pd attachment barrier by starting the growth of narrow Pd islands and then interrupting the Pd dosing for different periods of time. Stopping the deposition causes the growth to cease immediately. If we restart deposition three or more minutes later, none of the existing active regions resume growing. Instead, new islands are nucleated and new active step edges emerge from them, much as during the early state of the experiments. With shorter interruptions the original active regions of the step edges resume growth. This suggests that the alloy concentration necessary to block step edges takes some time to build up.

Can this time dependent alloy concentration profile near step edges account for the observed growth mode? When step edges are moving slowly, the alloy concentration should behave similarly to stationary steps: the alloy concentration next to the step edges increases until it reaches a relatively large (close-to-equilibrium) value of about 5% and then gradually extends onto the adjoining terraces. As shown schematically in Fig. 3(b), this situation leads to an instability in the growth front: the flux of Pd attaching at the step edge will be largest where the alloy concentration is (by random fluctuations, initially) the smallest. This increased attachment and higher step velocity causes the alloy concentration in front of the step to be even lower (because the time spent by the step in any one position is less) and Pd attachment to be even faster. If the speed of the steps becomes sufficiently large, one can envision a steady-state situation where the alloy concentration near the step edge remains at a constant low value below the equilibrium density. We propose that these slow and fast regimes are separated by a distinct critical speed. Below the critical speed, attachment at the Pd step edges becomes increasingly unfavorable with time as the alloy concentration builds up. Above the critical speed, the step edge moves faster than the speed at which the increased alloy concentration spreads onto the terrace, so that concentration remains low and Pd attachment stays easy.

The existence of such a critical velocity has direct experimental support. In Fig. 1 for example, no step propagates at a velocity less than 45 nm/s. Remarkably, the propagation speed is always very close to this minimum velocity, despite strong variations in the environment and size of the moving front. To rationalize this result we note that moving more quickly than the critical speed (given that the local flux remains constant) would result in a narrowing of the growth front (to conserve mass). This shrinking would be resisted by various effects. For example, a narrow growth front would require adatom concentration gradients perpendicular to the growth direction. These gradients would be unstable and would broaden (and thus slow) the growth front. The increased energy associated with large step curvatures would also oppose narrowing. Thus the growth front appears to propagate at the slowest possible sustainable speed [16].

In this scenario, the critical velocity should only depend on the rate of alloying but not on the overall deposition flux. This prediction is tested in LEEM experiments that use different Pd deposition rates. We have measured the growth velocity of deposition rates of 1/650, 1/1400, and 1/6300 ML/s. The measured active step velocities are 50, 44, and 30 nm/s, respectively. So, an order of magnitude change in flux leads only to a relatively small change in velocity, roughly consistent with our proposed model. Given approximately the same number of active regions per unit area, mass conservation then requires that the higher the flux, the wider the active regions. This effect is clearly observed in Fig. 4. Also the active regions tend to get narrower, but move at the same speed, at the late stages of growth when there are more active regions and the flux per active region is less. Notice that the flux dependence of the width rules out that the labyrinthine patterns can be explained in terms of equilibrium surface stress domains.

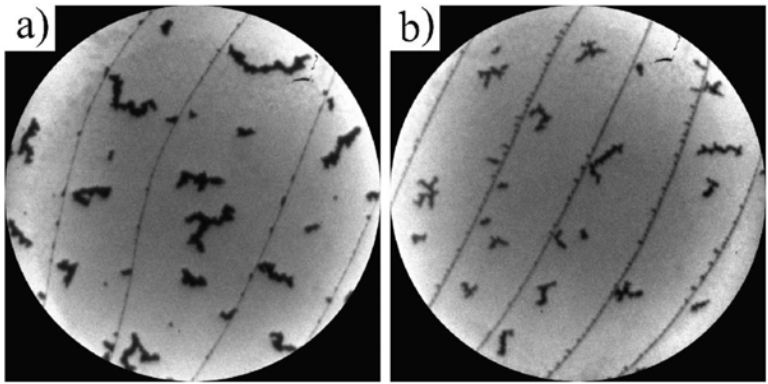

FIG. 4. LEEM images of Pd deposited on Ru(0001) at deposition rates and times of (a) 1/650 ML/s and 42 s and (b) 1/6300 ML/s and 190 s. The deposition temperature is 840 K; the field of view is 6 μm. The averaged speed of the advancing active regions is 50 nm/s and 30 nm/s, respectively.

Corroboration that surface alloying is responsible for the novel growth process comes by codepositing Ru with the Pd. According to the DFT calculations described above, one would expect the blocking Pd atoms embedded in the Ru substrate to be displaced by deposited Ru adatoms [17]. Consistent with this we find that even small simultaneous codeposition of Ru [with rates of 1/650 ML/s (Pd) and 1/1500 ML/s (Ru)] causes Pd to be uniformly incorporated into all step edges.

In summary, we have found strong evidence that surface alloying can impede step-flow growth during Pd/Ru epitaxy at around 840 K, leading to distinctive snakelike motion that gives rise to labyrinth patterns. The conclusion that an instability causes the observed growth mode might be applicable to other heteroepitaxial systems with surface alloying, in particular if the deposited species has weaker bonds and lower surface energy than the substrate species. In this general case the higher adatom formation energy of the substrate species should kinetically limit alloying on substrate terraces to regions close to step edges with all its consequences. The kinetic mechanism for this pattern formation is very different from the explanation of similar patterns in terms of surface stress domains, as in, e.g., Pb/Cu(111). It is plausible that labyrinthine growth also could occur in systems where impurity gas adsorption [18] creates step attachment barriers.

This research was partly supported by the Office of Basic Energy Sciences, Division of Materials Sciences, U.S. Department of Energy under Contracts No. DE-AC04-94AL85000 and No. DE-AC02-05CH11231, and by the Spanish Ministry of Science and Technology through Project No. MAT2006-13149-C02-02.